\documentclass[%
 reprint,
superscriptaddress,
bibnotes,
 amsmath,amssymb,
 aps,
prl,
]{revtex4-2}

\usepackage{graphicx}
\usepackage{dcolumn}
\usepackage{bm}
\usepackage{subfigure}
\usepackage{multirow}

\begin{document}

\preprint{APS/123-QED}

\title{Phase transition characteristics of Faraday waves}
\thanks{Peizhao Li and Tiancheng Yu contributed equally to this work}
\author{Peizhao Li}
\affiliation{
 School of Physics, Peking University, Beijing
 }
\author{Tiancheng Yu}
\affiliation{
 School of Physics, Peking University, Beijing
 }

\author{Xuechang Tu}
\affiliation{
 School of Electronics Engineering and Computer Science, Peking University, Beijing
}
\author{Han Yan}
\affiliation{
 School of Physics, Peking University, Beijing
 }
\author{Wei Wang}
\affiliation{
 School of Physics, Peking University, Beijing
 }
\author{Luqun Zhou}
\email{email: zhoulq@pku.edu.cn}
\affiliation{
 School of Physics, Peking University, Beijing
 }

\date{\today}

\begin{abstract}
Through experimentation, we have discovered that 
under varying driving conditions, the 
Faraday waves undergo two abrupt transitions in 
spatiotemporal order: onset and instability. 
The driving 
amplitudes and frequencies corresponding to these two 
transitions follow power-law relationships. 
The exponent of the power-law for the onset can be used to 
categorize different liquids into two distinct classes, 
which primarily reflects the differential contribution of 
surface tension and viscous forces in the surface wave 
dispersion relation. 
Furthermore, the exponent of the power-law for the 
instability serves as a quantitative indicator of the 
non-Newtonian properties of the liquid. Based on our 
experimental findings, we develop a phenomenological 
theoretical model that provides a comprehensive understanding 
of the properties of the Faraday pattern.
\end{abstract}

\maketitle

\section{Introduction}

The phenomenon of regular pattern formation on the surface of 
a fluid subjected to periodic vertical external forces, known 
as the Faraday pattern, was first discovered by M. Faraday in 
1831 \cite{faraday_xvii_1831}. Over the past few centuries, 
a multitude of experimental results have been reported, 
including pattern selection \cite{benjamin_stability_1954, zhang_pattern_1997,
binks_nonlinear_1997,wagner_faraday_1999,chen_amplitude_1999,
perinet_numerical_2009,skeldon_can_2015}, bifurcation 
\cite{douady_experimental_1990}, oscillon 
\cite{arbell_temporally_2000,xia_oscillon_2012,
chen_identifying_2018}, lattice defects 
\cite{ezersky_dynamics_1995}, and turbulence 
\cite{tufillaro_order-disorder_1989,wright_diffusing_1996,
shani_localized_2010,von_kameke_double_2011,francois_three-dimensional_2014}. 
These extensive investigations have greatly enriched our 
understanding of the Faraday waves and even nonlinear systems.

For a system undergoing a transition from order to disorder, 
undoubtedly, the transition parameters that correspond to 
its phase transition point and the behavior of the system 
in the vicinity of the phase transition point are the two 
fundamental issues that merit our attention. The pertinent 
research was initially conducted by Tufillaro et al. in 1989 \cite{tufillaro_order-disorder_1989}, where they considered the spatial auto-correlation function of Faraday waves and obtained the phase transition curve. Subsequent studies have further augmented and refined the system's phase diagram on the one hand \cite{wagner_faraday_1999}, and explored the mechanism of the phase transition using different methods on the other hand \cite{shani_localized_2010,francois_three-dimensional_2014}. These studies have furnished us with a comprehensive and lucid portrayal of the Faraday wave phase transition.

In this article, we approach the Faraday pattern transitions 
from a different perspective, focusing on the relationship 
between the external driving amplitudes and frequencies. 
Through experiments, we discover power-law relationships 
between the transition driving acceleration and frequency at 
the onset of the Faraday waves. Based on 
this power-law exponent, we define a characteristic wave 
number of the fluid surface wave to characterize its 
dynamical properties. Furthermore, we find that with a 
further increase in the driving amplitude, the Faraday 
waves also undergo a stable-to-unstable transition, and 
the transition acceleration and frequency also follow a 
power-law relationship, which may reflect the non-Newtonian 
nature of the fluid. Based on experimental data, 
we develop a phenomenological theoretical model that 
provides a unified understanding of the Faraday pattern 
properties.

\section{Methods}
\begin{figure}
\centering
\includegraphics[width = 7cm]{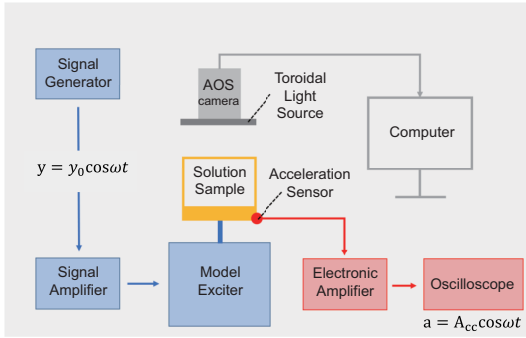}
\caption{Diagram of the experimental setup}
\label{equipment}
\end{figure}

The experimental setup employed in this study is 
illustrated in FIG. \ref{equipment}, which consists of a 
cylindrical container with a diameter of 14 cm, 
filled with a liquid sample that is 2 cm thick. 

The excitation signal generated by the 
signal generator is amplified by the 
power amplifier to drive the modal 
exciter (Sinocera Piezotronics, JZK-100), causing the liquid sample 
to vibrate sinusoidally at different 
amplitudes and frequencies. Meanwhile, 
the oscilloscope displays the accelerometer 
signal amplified by the charge amplifier in 
real-time.

To capture the shape 
of the liquid surface Faraday waves, we 
employ a high-speed camera (AOS Technologies AG, S-PRI F2) 
to record the 
reflection of light on the liquid surface, 
which is further enhanced with the addition 
of a drop of ink to increase the image contrast. 
The entire experimental setup is maintained 
at a constant temperature of 26$^{\circ}$C.

\begin{figure}
\centering
\includegraphics[width = 7cm]{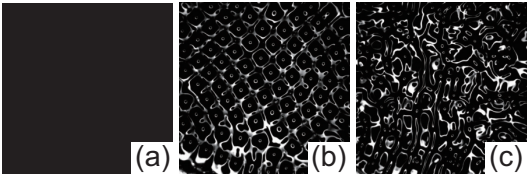}
\caption{Surface patterns of 5g/L xanthan gum fluid under different driving conditions: (a) no Faraday waves ($f = 60\rm{Hz}$, $A_{cc} = 3.92\rm{g}$); (b) ordered Faraday waves ($f = 60\rm Hz$, $A_{cc}=5.77$g); (c) disordered Faraday waves ($f = 60\rm Hz$, $A_{cc}=7.00$g).}
\label{photo}
\end{figure}
The 
behaviors of the Faraday waves exhibit distinct phase 
transitions: at low driving amplitudes, the liquid moves uniformly with 
the container, and the surface displays no observable pattern 
(FIG. \ref{photo}a). Beyond a threshold $A_{c1}$, the surface exhibits 
the spontaneous emergence of isolated oscillators, which 
gradually stabilize in amplitude (FIG. \ref{photo}b). Increasing the 
driving amplitude further causes the oscillators to persist 
independently until reaching another threshold $A_{c2}$, 
beyond which they collide and merge, resulting in a 
disordered surface pattern (FIG. \ref{photo}c). We term the transition 
from no pattern to pattern as the "onset," and the transition 
from ordered to disordered as "instability" for clarity.

We can easily determine the transition point at 
which the onset of Faraday waves occurs by visual inspection, 
while for the transition point at which the Faraday waves 
become unstable,
we use two methods to measure the degree of order in 
both temporal and spatial dimensions of the data obtained 
from the experiments.

In the temporal dimension, we select a frame as the 
starting point from the time sequence images obtained from 
the experiments and calculate the autocorrelation function 
between subsequent frames and the starting point:
\begin{equation}
r(t) = \frac{\sum_{i,j}(P_{ij}(t)-\bar{P}(t))(P_{ij}(0)-\bar{P}(0))}{\sqrt{\sum_{i,j}(P_{ij}(t)-\bar{P}(t))^2(P_{ij}(0)-\bar{P}(0))^2}}
\end{equation}
where $P_{ij}$ represents the gray value of the $(i,j)$ 
pixel in the image. Furthermore, for the case where 
the driving period is $T$ and the acceleration amplitude 
is $A_{cc}$, we calculate the reproducibility:
\begin{equation}
F(T,A_{cc}) = \frac{r(T)+r(2T)+r(3T)}{3}
\end{equation}

Since the probability of two frames being accidentally correlated in the time sequence images is very low (about $10^{-7}$), 
the reproducibility can well reflect the temporal periodicity of the system.

\begin{figure}
\centering
\includegraphics[width=7cm]{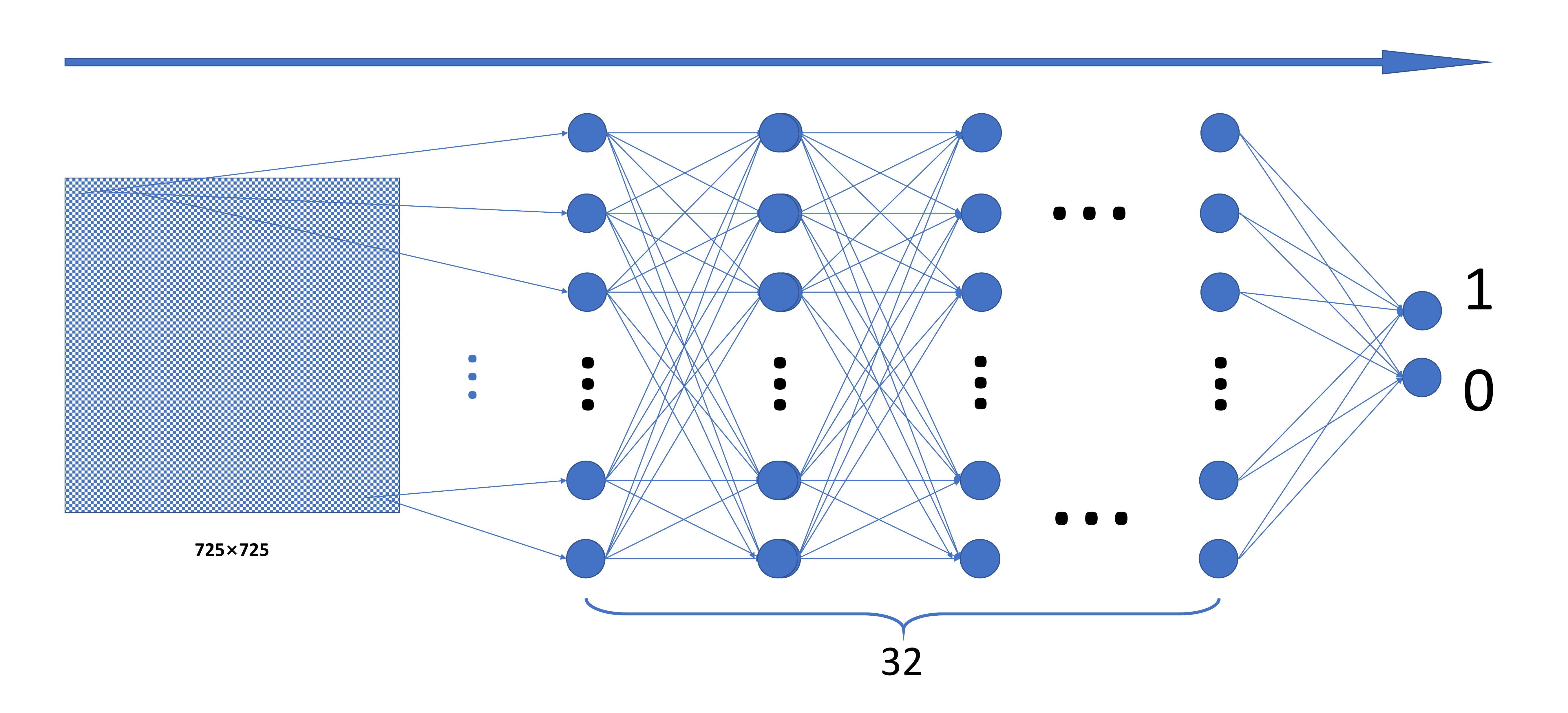}
\caption{Schematic diagram of clustering algotithm}
\label{al}
\end{figure}

In the spatial dimension,
we employ a clustering algorithm (FIG. \ref{al}) to 
categorize images of the liquid surface into ordered and 
disordered classes in the spatial dimension, using the 
confidence probability $p$ of each image's classification 
as a measure of its orderliness. To ensure a representative 
image set and reduce extraneous factors, we exclude the top 
15\% and bottom 30\% of images with the highest and lowest 
orderliness, respectively, and calculate the average degree 
of order for the remaining images as the spatial orderliness 
$q$. The algorithm uses distribution features to classify 
the images, and if a small change in parameters results in 
most images switching classes, we conclude that the Faraday 
wave's spatial characteristics have changed.

\section{results}
Our experiments reveal two phase transitions in the 
Faraday pattern: the onset and instability transitions. 
We observe a significant power-law relationship between 
the driving amplitude and frequency of both transitions, 
with corresponding exponents summarized in TABLE I. Xanthan 
gum at different concentrations exhibit onset transition 
exponents around 3/2, while silicone oil and various 
glycerol concentrations display transition exponents of 
approximately 5/3, consistent with previous studies on CO$_2$ 
liquid-gas interface\cite{fauve_parametric_1992} and 
paraffin oil\cite{ezersky_dynamics_1995} (TABLE I). 
For Newtonian fluids, the instability transition exponent 
remains constant and close to 5/3 regardless of viscosity. 
Conversely, for the same non-Newtonian fluid, the transition 
exponent decreases linearly with increasing viscosity. 
We will provide a detailed analysis of our experimental 
findings in subsequent sections.

\begin{table*}
\centering
\renewcommand\arraystretch{1.2} 
\begin{tabular}{|c|c|c|c|c|}
\hline
Liquid Type                   & Liquid          & viscocity/Pa$\cdot$s & Onset Transition Exponent   & Instability Transition Exponent   \\ \hline
\multirow{6}{*}{Newtonian fluids} & CO$_2$ liquid-gas interface\cite{fauve_parametric_1992}&&1.76& \\ \cline{2-5}
&Paraffin oil\cite{ezersky_dynamics_1995}&0.05&1.70& \\ \cline{2-5}
						  & Silicone oil          & 0.35    & $1.67 \pm 0.03$ & $1.75 \pm 0.04$ \\ \cline{2-5} 
                       & Glycerin          & 1.2    & $1.75 \pm 0.03$ & $1.64 \pm 0.02$ \\ \cline{2-5} 
                       & 85\%Glycerin      & 0.13    & $1.67 \pm 0.03$ & $1.64 \pm 0.01$ \\ \cline{2-5} 
                       & 50\%Glycerin      & 0.05      & $1.75 \pm 0.03$ & $1.59 \pm 0.03$ \\ \hline
\multirow{4}{*}{Non-Newtonian fluids} & 2.5g/L Xanthan gum solution & 20   & $1.52\pm 0.03$ & $1.78 \pm 0.02$\\ \cline{2-5} 
                       & 5g/L Xanthan gum solution   & 30  & $1.48 \pm 0.02$ & $1.51 \pm 0.02$\\ \cline{2-5} 
                       & 10g/L Xanthan gum solution  & 70   & $1.49 \pm 0.02$ & $1.34 \pm 0.01$\\ \cline{2-5} 
                       & 15g/L Xanthan gum solution  & 80   & $1.49 \pm 0.01$ & $1.24 \pm 0.03$\\ \hline
\end{tabular}

\caption{Summary table of onset and instability transition indices for fluid Faraday waves.
}
\end{table*}

\subsection{Phase transition characteristics of Faraday waves}

\begin{figure*}
\centering
\includegraphics[width=15cm]{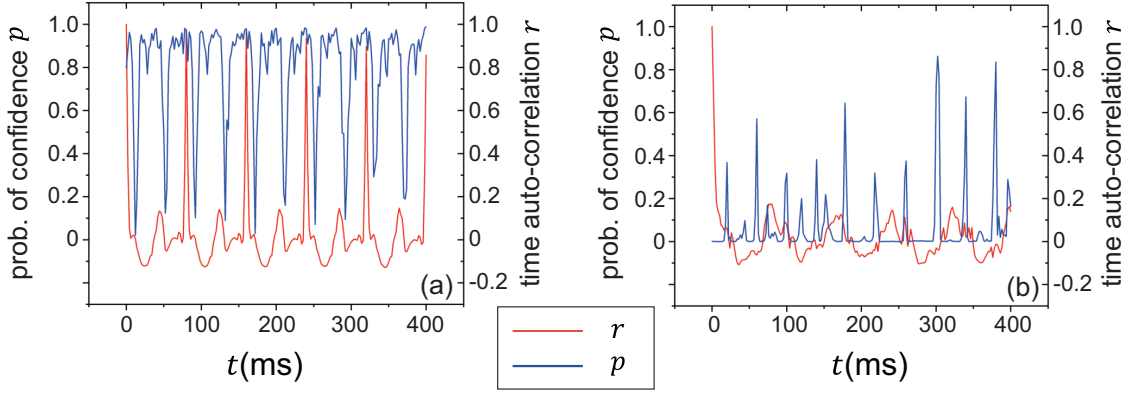}
\caption{The figure above shows the plots of the time autocorrelation function $r$ (in red) and spatially ordered confidence probability $p$ (in blue) of 5g/L xanthan gum solution under different driving conditions. (a) $f=25$Hz, $A_{cc}=1.83$g; (b) $f=25$Hz, $A_{cc}=2.34$g.}
\label{time}
\end{figure*}

We observe the surface patterns of different liquids 
under various driving conditions. 
To locate the transition point at which the Faraday waves 
become unstable, 
we take a 5 g/L solution of xanthan gum as 
an example and 
plot the temporal variation of the two indicators $r$ and $p$ 
under different 
driving conditions with a frequency of $f=25$Hz 
in FIG. \ref{time}. 

At low driving amplitudes (FIG. \ref{time}a), 
$r$ varies periodically with time at a frequency identical 
to that of the driving force, while $p$ remains close to 1 
in most cases and only periodically exhibits low values at 
a frequency that is twice the driving force frequency. 
This is because, for a solitary wave in an ordered state, 
the shape of the liquid surface is mainly determined by the 
disordered noise when it vibrates to the equilibrium position. 
As the driving amplitude increases (FIG. \ref{time}b), 
$r$ rapidly drops from 1 to near 0, and $p$ remains mostly 
low, only occasionally exhibiting high values when the 
liquid surface forms more ordered structures.

\begin{figure}
\centering
\includegraphics[width=7cm]{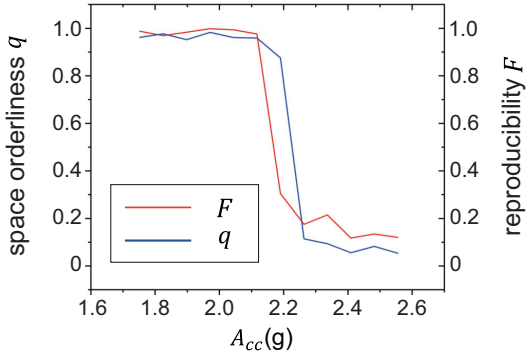}
\caption{Graph of the variation of the time reproducibility $F$ and spatial order degree $q$ of 5g/L xanthan gum solution with respect to the amplitude $A_{cc}$ at a driving frequency of 25Hz.}
\label{transition}
\end{figure}

To further analyze the behavior of the 5g/L xanthan gum 
solution, we calculated the time reproducibility $F$ and 
space orderliness $q$ based on the $r$ and $p$ values 
obtained under different driving conditions. 
The variations of $F$ and $q$ with the driving amplitude 
$A_{cc}$ under a driving frequency of $f=25$Hz are shown 
in FIG. \ref{transition}. As the driving amplitude increases, 
both $F$ and $q$ initially remain close to 1, but once a 
transition value is reached, they suddenly drop to nearly 0, 
corresponding to a phase transition from order to disorder 
in both time and space. As can be seen from the FIG. \ref{transition}, 
the transition position where the time reproducibility 
represented by $F$ and the spatial orderliness represented 
by $p$ undergo a phase transition is within the range of 
the measurement sampling interval.
Due to the essentially coincident time-space phase 
transition point, we no longer differentiate between the two, 
and instead use the value of $A_{c2}$, corresponding to when 
$F$ drops to 0.5, as the instability phase transition point 
of the pattern.

\subsection{The power-law relationship between the transition amplitudes and frequencies}
\begin{figure}
\centering
\includegraphics[width=7cm]{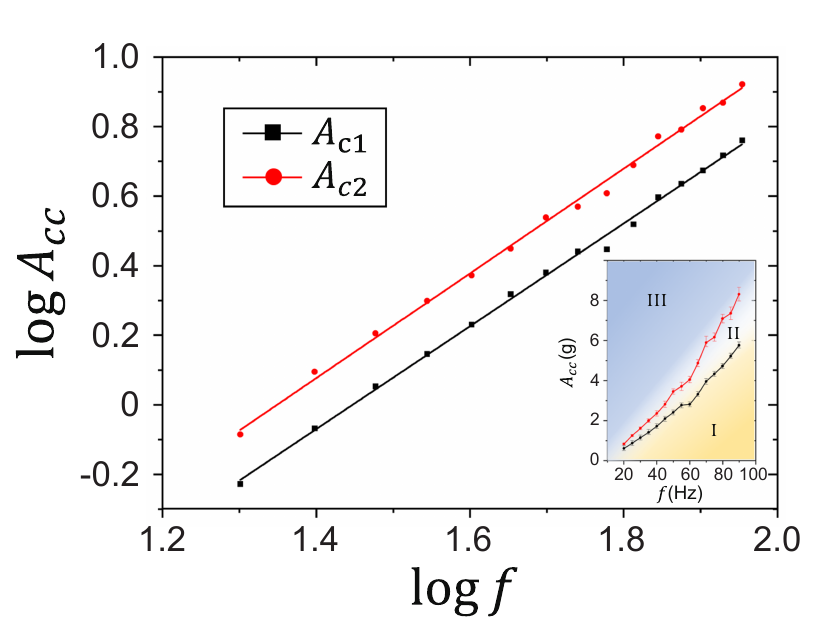}
\caption{log$A$-log$f$ linear fitting graph 
of 5g/L xanthan gum fluid. Transition onset amplitude $A_{c1}\propto f^{1.48}$,  corresponding to $r=0.9987$; instability amplitude $A_{c2}\propto f^{1.51}$, corresponding to $r = 0.9983$. (Insert FIG.) Faraday waves phase diagram. Region I: No pattern; Region II: Stable pattern; Region III: Unstable pattern.}
\label{phase}
\end{figure}

By utilizing the method described in the previous section, 
we are able to experimentally determine the onset transition 
amplitude $A_{c1}$ and the instability transition amplitude 
$A_{c2}$ under a fixed driving frequency $f$. 
Taking 5g/L of xanthan gum as an example, 
we plot the $A_{c1}$ and $A_{c2}$ at different $f$ together 
to construct the phase diagram of the system, 
as illustrated in the inset of FIG. \ref{phase}. 
The phase diagram divides the driving parameter space 
into three regions: region I where no pattern is formed, 
region II where stable patterns emerge, and region III 
where unstable patterns arise.

Furthermore, we apply a double logarithmic scale to fit 
the relationship between $A_c$ and $f$ and 
a strong power-law correlation between the two. 
The power-law exponents for $A_{c1}$ and $A_{c2}$ are 
determined to be 1.48 and 1.51, respectively. 
This finding raises an important question: 
if $A_c \propto f$ or $A_c \propto f^2$ at the transition 
point, it implies that the transition phenomenon is 
affected by the driving speed or amplitude from the external 
environment. However, our experimental data clearly indicates 
that $A_c \propto f^{1.5}$, indicating that this transition 
exponent is determined by intrinsic properties of the 
system's dynamics. In subsequent sections, we analyze these 
two transition exponents independently.

\subsection{The onset transition exponent}
We first focus on the transition amplitude $A_{c1}$ for the 
onset of oscillations. To investigate this further, 
we conduct the same experiment using various concentrations 
of xanthan gum, glycerol-water mixtures, and silicone oil. 
We plot the double-logarithmic fitting curves for $A_{c1}$ 
versus $f$ and compile their slopes in TABLE I. 
Despite the different transition amplitudes required for the 
onset of oscillations at the same driving frequency for each 
liquid, the power-law relationship between $A_{c1}$ and $f$ 
remains consistent and the power-law exponents 
exhibit distinct distribution characteristics. 
Specifically, different concentrations of xanthan gum exhibit 
power-law exponents around 3/2, whereas silicone oil and 
different concentrations of glycerol exhibit power-law 
exponents around 5/3.

We further analyze this result theoretically:

Based on the findings by S. Fauve et al. in 1992 \cite{fauve_parametric_1992}, in the absence of air density and viscosity, a transition amplitude $a_{c1}$ for the 
Faraday pattern to appear is related to the driving frequency $f$ by:
\begin{equation}
A_{c1} \approx 8\pi fk \frac{\nu}{\rho}
\end{equation}
where $k$ is the wave vector of the liquid surface waves and $\nu$ is the dynamic viscosity of the liquid. For viscous fluids, the surface shallow water waves satisfy the dispersion relation \cite{zhang_pattern_1997}:
\begin{equation}
\omega^2=[gk+\frac{\alpha}{\rho}k^3+4(\frac{\nu}{\rho})^2k^4]\tanh(Hk)
\end{equation} 

Here, $g$ is the acceleration due to gravity, 
$\alpha$ is the liquid surface tension coefficient, 
$\rho$ is the liquid density, and $H$ is the liquid depth. 
In our experiment, $k\sim 400-1000\rm m^{-1}$, $Hk \gg 1$, 
and $\tanh(Hk) \approx 1$, 
while $gk \ll \frac{\alpha}{\rho}k^3, (\frac{\nu}{\rho})^2k^4$. Thus, 
the dispersion is mainly determined by the cubic and 
quartic terms of $k$, and the ratio of the two terms is:
\begin{equation}
\eta = \frac{w^2_{k^4}}{w^2_{k^3}}=\frac{4\nu^2 k}{\rho \alpha}
\end{equation}

For different concentrations of xanthan gum solution
($\nu \sim 10 \rm Pa\cdot s$, $\rho \sim 10^3 \rm kg/m^3$,
$\alpha \sim 10^{-2} \rm N/m$), $\eta \sim 10^4$, the dispersion relation is
mainly determined by the quartic term $k^4$, resulting in
$k \propto \omega^{1/2}$ and $a_{c1} \propto f^{3/2}$, which
is consistent with our experimental observations.
Similarly, for silicon oil and glycerol-water solutions
($\nu \sim 10^{-2} \rm Pa\cdot s$, $\rho \sim 10^3 \rm kg/m^3$,
$\alpha \sim 10^{-2} \rm N/m$), $\eta \sim 10^{-2}$ the dispersion relation is mainly
determined by the cubic term $k^3$, leading to $k \propto f^{2/3}$
and $a_{c1} \propto f^{5/3}$.These theoretical results are in good agreement with 
our experimental observations.

We can extract a characteristic wave number $k_c$ from Eq. (5) to characterize the dynamic properties of the Faraday pattern:
\begin{equation}
k_c = \frac{\rho\alpha}{4\nu^2}
\end{equation}

The relationship between $a_{c1}$ and $f$ depends on the 
relative magnitude of $k_c$ and $k$: when $k_c \ll k$, 
$a_{c1} \propto f^{3/2}$; when $k_c \gg k$, $a_{c1} \propto 
f^{5/3}$.

\subsection{The instability transition exponent}
Next, we analyze the transition amplitude for instability, 
$A_{c2}$. As shown in TABLE I, the transition exponent for 
instability remains nearly constant and close to 5/3 for 
Newtonian fluids, irrespective of their viscosity. 
Conversely, for non-Newtonian fluids, the transition 
exponent decreases linearly with increasing viscosity.

We provide a semi-quantitative explanation of this result:

Starting from the Navier-Stokes equations for an ideal fluid
incorporating surface tension, we find the gradient for it and obtain the generalized
equations by adding the viscous term and generalizing
the acceleration term:
\begin{equation}
        \frac{\partial v}{\partial t}=g^*+\frac{\alpha}{\rho}\nabla(\frac{\partial^2 \zeta}{\partial x^2}+\frac{\partial^2 \zeta}{\partial y^2})+\frac{\nu}{\rho}\Delta v
   \end{equation}
where $g^*$ 
representing the effective gravitational acceleration 
in the platform's reference frame.
For the case involved in this 
experiment, we can write:
\begin{equation}
        g^*=g+A_{cc} \cos{(2\pi ft+\phi)}
\end{equation}
where $g$ and $A_{cc}$ are constant vectors.\par
We calculate the scalar product of $v$ on both sides of Eq. (7) and average it over the spatial region of oscillation. By applying the scattering theorem (which states that the scattering of the velocity field of an incompressible fluid is zero), the spatial integral of the second term on the right side of Eq. (7) can be reduced to an integral with zero results at the region boundary. We can handle the last term using the divisional integration approach:
\begin{equation}
        \frac{1}{2} \frac{\partial \langle |v|^2 \rangle}{\partial t}=\langle vg^* \rangle-\frac{\nu}{\rho} \langle \nabla v:\nabla v \rangle    
    \end{equation}
where $\nabla v:\nabla v=(\partial_k v_i)(\partial_k v_i)$.\par
As the system is subjected to steady periodic vibrations, the time derivative of the spatial average in the left-hand side equation is zero. The subscript $m$ is used to denote the amplitude of each vibrational quantity, and hence the following approximate expression can be obtained:
\begin{equation}
        v_m A_{cc} \propto \frac{\nu}{\rho} k^2 {v_m}^2
\end{equation}
where $k$ is the wave number of surface tension gravity waves. From the energy point of view, this equation can be approximated as the energy balance equation, i.e. the work done by the external "gravitational field" is balanced by the energy dissipated by the viscous dissipation of the oscillating fluid.\par

If we denote the amplitude of the oscillator as 
$A_m$, then we have $v_m \sim A_mf$. Substituting this 
expression into Eq. (10), we obtain:
\begin{equation}
A_{cc}\sim \frac{\nu}{\rho}k^2fA_m
\end{equation}

The instability of Faraday waves is attributed to the interaction between neighboring oscillators, which occurs when the diameter of a single oscillator exceeds the wavelength $\lambda$. This interaction causes the originally independent oscillators to become coupled and leads to instability. 
Moreover, the shape of the oscillators is nearly constant, which means $d \propto A_m$. When the pattern becomes unstable, $A_m \sim 1/k$, that is:

\begin{equation}
A_{c2}\sim \frac{\nu}{\rho}kf
\end{equation}

This transition driving amplitude $A_{c2}$ corresponds to the 
Faraday wave pattern and has a similar form to Eq. (3).

For the Newtonian fluid used in the experiment with 
relatively low viscosity, its viscosity remains constant, 
and the wavenumber is proportional to $f^{2/3}$. 
Substituting this into Eq. (12), we obtain
\begin{equation}
A_{c2}\propto f^{5/3}
\end{equation}

For non-Newtonian fluids, the relationship between viscosity and oscillation frequency follows\cite{cross_relation_1979}:
\begin{equation}
\nu = \nu_0 f^{n-1}
\end{equation}
where $n<1$ and decreases with increasing $\nu_0$, indicating a more significant shear-thinning effect in fluids with higher initial viscosity. In the experiment, the xanthan gum solution experiences a drop in viscosity to $10^{-2}\rm Pa \cdot s$ due to the shear-thinning effect when it becomes unstable. Therefore, its dispersion relation can be approximated as $k\propto f^{2/3}$. 
Substituting this into Eq. (12) gives:
\begin{equation}
A_{c2}\propto f^{n+2/3}
\end{equation}

The power-law exponent $n$ and $\nu_0$ for the xanthan gum solution can be related through a fitted curve obtained from the experimental data, given by $n=1.07-0.0075\nu_0$. This suggests that the shear-thinning behavior is absent when the viscosity is zero, as $n\approx 1$. As the viscosity increases, the shear-thinning effect becomes more prominent, which agrees with the experimental findings.

It should be noted that in the preceding section, the shear-thinning effect was not taken into account when determining the driving amplitude for the onset of oscillations. This is because, prior to the onset of oscillations, there is no shear rate at the liquid surface and the liquid's dynamic properties are governed only by its intrinsic viscosity $\nu_0$. However, the shear rate at the liquid surface is significant before and after the instability, making the shear-thinning effect important and cannot be neglected.

\section{Conclusion}
In this study, we have conducted experiments to investigate the temporal reproducibility and spatial orderliness of Faraday waves in different liquids under various driving conditions. We have discovered a discontinuous change in the degree of spatial orderliness and temporal reproducibility of the waves at the onset of instability, and we have found that the transition driving amplitude and frequency at this point follow a power-law relationship. The transition exponents for the onset of wave patterns have been found to be 3/2 and 5/3 for two groups of liquids, respectively, which depend on the relative importance of the surface tension and viscosity terms in the shallow-water wave dispersion relation. We have also proposed a characteristic wave number, $k_c$, that characterizes the dynamical properties of Faraday waves.

Regarding the instability of Faraday waves, we have observed that the transition exponents of Newtonian fluids are near 5/3, while those of non-Newtonian fluids decrease almost linearly with increasing viscosity. We have hypothesized that the instability of the pattern originates from the interaction between neighboring oscillators after the increase in their size. Through a semi-quantitative analysis, we have provided a theoretical explanation for the transition exponent of Newtonian fluids and proposed that the shear-thinning effect may cause the transition exponent of non-Newtonian fluids to vary with viscosity. 

Restricted by experimental conditions, our study is limited to the cases of $\eta \gg 1$ and $\eta \ll 1$, where $\eta$ is the dimensionless viscosity, and further experiments are needed to verify the findings for the case of $\eta\sim 1$, where the contribution of the surface tension and viscosity terms to the shallow-water wave dispersion relation is of similar magnitude. Additionally, the understanding of the instability of non-Newtonian fluids requires further theoretical exploration.

The work is supported by the Undergraduate 
Teaching Reform Project at Peking University(JG2023003). 
We appreciate Prof. Qi Ouyang for many useful discussions. 


\bibliography{apssamp}

\begin{thebibliography}{20}%
\makeatletter
\providecommand \@ifxundefined [1]{%
 \@ifx{#1\undefined}
}%
\providecommand \@ifnum [1]{%
 \ifnum #1\expandafter \@firstoftwo
 \else \expandafter \@secondoftwo
 \fi
}%
\providecommand \@ifx [1]{%
 \ifx #1\expandafter \@firstoftwo
 \else \expandafter \@secondoftwo
 \fi
}%
\providecommand \natexlab [1]{#1}%
\providecommand \enquote  [1]{``#1''}%
\providecommand \bibnamefont  [1]{#1}%
\providecommand \bibfnamefont [1]{#1}%
\providecommand \citenamefont [1]{#1}%
\providecommand \href@noop [0]{\@secondoftwo}%
\providecommand \href [0]{\begingroup \@sanitize@url \@href}%
\providecommand \@href[1]{\@@startlink{#1}\@@href}%
\providecommand \@@href[1]{\endgroup#1\@@endlink}%
\providecommand \@sanitize@url [0]{\catcode `\\12\catcode `\$12\catcode
  `\&12\catcode `\#12\catcode `\^12\catcode `\_12\catcode `\%12\relax}%
\providecommand \@@startlink[1]{}%
\providecommand \@@endlink[0]{}%
\providecommand \url  [0]{\begingroup\@sanitize@url \@url }%
\providecommand \@url [1]{\endgroup\@href {#1}{\urlprefix }}%
\providecommand \urlprefix  [0]{URL }%
\providecommand \Eprint [0]{\href }%
\providecommand \doibase [0]{https://doi.org/}%
\providecommand \selectlanguage [0]{\@gobble}%
\providecommand \bibinfo  [0]{\@secondoftwo}%
\providecommand \bibfield  [0]{\@secondoftwo}%
\providecommand \translation [1]{[#1]}%
\providecommand \BibitemOpen [0]{}%
\providecommand \bibitemStop [0]{}%
\providecommand \bibitemNoStop [0]{.\EOS\space}%
\providecommand \EOS [0]{\spacefactor3000\relax}%
\providecommand \BibitemShut  [1]{\csname bibitem#1\endcsname}%
\let\auto@bib@innerbib\@empty
\bibitem [{\citenamefont {Faraday}(1831)}]{faraday_xvii_1831}%
  \BibitemOpen
  \bibfield  {author} {\bibinfo {author} {\bibfnamefont {M.}~\bibnamefont
  {Faraday}},\ }\href {https://doi.org/10.1098/rstl.1831.0018} {\bibfield
  {journal} {\bibinfo  {journal} {Philosophical Transactions of the Royal
  Society of London}\ }\textbf {\bibinfo {volume} {121}},\ \bibinfo {pages}
  {299} (\bibinfo {year} {1831})},\ \bibinfo {note} {publisher: Royal
  Society}\BibitemShut {NoStop}%
\bibitem [{\citenamefont {Benjamin}\ \emph {et~al.}(1954)\citenamefont
  {Benjamin}, \citenamefont {Ursell},\ and\ \citenamefont
  {Taylor}}]{benjamin_stability_1954}%
  \BibitemOpen
  \bibfield  {author} {\bibinfo {author} {\bibfnamefont {T.~B.}\ \bibnamefont
  {Benjamin}}, \bibinfo {author} {\bibfnamefont {F.~J.}\ \bibnamefont
  {Ursell}},\ and\ \bibinfo {author} {\bibfnamefont {G.~I.}\ \bibnamefont
  {Taylor}},\ }\href {https://doi.org/10.1098/rspa.1954.0218} {\bibfield
  {journal} {\bibinfo  {journal} {Proceedings of the Royal Society of London.
  Series A. Mathematical and Physical Sciences}\ }\textbf {\bibinfo {volume}
  {225}},\ \bibinfo {pages} {505} (\bibinfo {year} {1954})},\ \bibinfo {note}
  {publisher: Royal Society}\BibitemShut {NoStop}%
\bibitem [{\citenamefont {Zhang}\ and\ \citenamefont
  {Viñals}(1997)}]{zhang_pattern_1997}%
  \BibitemOpen
  \bibfield  {author} {\bibinfo {author} {\bibfnamefont {W.}~\bibnamefont
  {Zhang}}\ and\ \bibinfo {author} {\bibfnamefont {J.}~\bibnamefont
  {Viñals}},\ }\href {https://doi.org/10.1017/S0022112096004764} {\bibfield
  {journal} {\bibinfo  {journal} {Journal of Fluid Mechanics}\ }\textbf
  {\bibinfo {volume} {336}},\ \bibinfo {pages} {301} (\bibinfo {year}
  {1997})},\ \bibinfo {note} {publisher: Cambridge University
  Press}\BibitemShut {NoStop}%
\bibitem [{\citenamefont {Binks}\ and\ \citenamefont {van~de
  Water}(1997)}]{binks_nonlinear_1997}%
  \BibitemOpen
  \bibfield  {author} {\bibinfo {author} {\bibfnamefont {D.}~\bibnamefont
  {Binks}}\ and\ \bibinfo {author} {\bibfnamefont {W.}~\bibnamefont {van~de
  Water}},\ }\href {https://doi.org/10.1103/PhysRevLett.78.4043} {\bibfield
  {journal} {\bibinfo  {journal} {Physical Review Letters}\ }\textbf {\bibinfo
  {volume} {78}},\ \bibinfo {pages} {4043} (\bibinfo {year} {1997})},\ \bibinfo
  {note} {publisher: American Physical Society}\BibitemShut {NoStop}%
\bibitem [{\citenamefont {Wagner}\ \emph {et~al.}(1999)\citenamefont {Wagner},
  \citenamefont {Müller},\ and\ \citenamefont {Knorr}}]{wagner_faraday_1999}%
  \BibitemOpen
  \bibfield  {author} {\bibinfo {author} {\bibfnamefont {C.}~\bibnamefont
  {Wagner}}, \bibinfo {author} {\bibfnamefont {H.~W.}\ \bibnamefont
  {Müller}},\ and\ \bibinfo {author} {\bibfnamefont {K.}~\bibnamefont
  {Knorr}},\ }\href {https://doi.org/10.1103/PhysRevLett.83.308} {\bibfield
  {journal} {\bibinfo  {journal} {Physical Review Letters}\ }\textbf {\bibinfo
  {volume} {83}},\ \bibinfo {pages} {308} (\bibinfo {year} {1999})}\BibitemShut
  {NoStop}%
\bibitem [{\citenamefont {Chen}\ and\ \citenamefont
  {Viñals}(1999)}]{chen_amplitude_1999}%
  \BibitemOpen
  \bibfield  {author} {\bibinfo {author} {\bibfnamefont {P.}~\bibnamefont
  {Chen}}\ and\ \bibinfo {author} {\bibfnamefont {J.}~\bibnamefont {Viñals}},\
  }\href {https://doi.org/10.1103/PhysRevE.60.559} {\bibfield  {journal}
  {\bibinfo  {journal} {Physical Review E}\ }\textbf {\bibinfo {volume} {60}},\
  \bibinfo {pages} {559} (\bibinfo {year} {1999})},\ \bibinfo {note}
  {publisher: American Physical Society}\BibitemShut {NoStop}%
\bibitem [{\citenamefont {Périnet}\ \emph {et~al.}(2009)\citenamefont
  {Périnet}, \citenamefont {Juric},\ and\ \citenamefont
  {Tuckerman}}]{perinet_numerical_2009}%
  \BibitemOpen
  \bibfield  {author} {\bibinfo {author} {\bibfnamefont {N.}~\bibnamefont
  {Périnet}}, \bibinfo {author} {\bibfnamefont {D.}~\bibnamefont {Juric}},\
  and\ \bibinfo {author} {\bibfnamefont {L.~S.}\ \bibnamefont {Tuckerman}},\
  }\href {https://doi.org/10.1017/S0022112009007551} {\bibfield  {journal}
  {\bibinfo  {journal} {Journal of Fluid Mechanics}\ }\textbf {\bibinfo
  {volume} {635}},\ \bibinfo {pages} {1} (\bibinfo {year} {2009})},\ \bibinfo
  {note} {publisher: Cambridge University Press}\BibitemShut {NoStop}%
\bibitem [{\citenamefont {Skeldon}\ and\ \citenamefont
  {Rucklidge}(2015)}]{skeldon_can_2015}%
  \BibitemOpen
  \bibfield  {author} {\bibinfo {author} {\bibfnamefont {A.~C.}\ \bibnamefont
  {Skeldon}}\ and\ \bibinfo {author} {\bibfnamefont {A.~M.}\ \bibnamefont
  {Rucklidge}},\ }\href {https://doi.org/10.1017/jfm.2015.388} {\bibfield
  {journal} {\bibinfo  {journal} {Journal of Fluid Mechanics}\ }\textbf
  {\bibinfo {volume} {777}},\ \bibinfo {pages} {604} (\bibinfo {year}
  {2015})},\ \bibinfo {note} {publisher: Cambridge University
  Press}\BibitemShut {NoStop}%
\bibitem [{\citenamefont {Douady}(1990)}]{douady_experimental_1990}%
  \BibitemOpen
  \bibfield  {author} {\bibinfo {author} {\bibfnamefont {S.}~\bibnamefont
  {Douady}},\ }\href {https://doi.org/10.1017/S0022112090003603} {\bibfield
  {journal} {\bibinfo  {journal} {Journal of Fluid Mechanics}\ }\textbf
  {\bibinfo {volume} {221}},\ \bibinfo {pages} {383} (\bibinfo {year}
  {1990})},\ \bibinfo {note} {publisher: Cambridge University
  Press}\BibitemShut {NoStop}%
\bibitem [{\citenamefont {Arbell}\ and\ \citenamefont
  {Fineberg}(2000)}]{arbell_temporally_2000}%
  \BibitemOpen
  \bibfield  {author} {\bibinfo {author} {\bibfnamefont {H.}~\bibnamefont
  {Arbell}}\ and\ \bibinfo {author} {\bibfnamefont {J.}~\bibnamefont
  {Fineberg}},\ }\href {https://doi.org/10.1103/PhysRevLett.85.756} {\bibfield
  {journal} {\bibinfo  {journal} {Physical Review Letters}\ }\textbf {\bibinfo
  {volume} {85}},\ \bibinfo {pages} {756} (\bibinfo {year} {2000})}\BibitemShut
  {NoStop}%
\bibitem [{\citenamefont {Xia}\ \emph {et~al.}(2012)\citenamefont {Xia},
  \citenamefont {Maimbourg}, \citenamefont {Punzmann},\ and\ \citenamefont
  {Shats}}]{xia_oscillon_2012}%
  \BibitemOpen
  \bibfield  {author} {\bibinfo {author} {\bibfnamefont {H.}~\bibnamefont
  {Xia}}, \bibinfo {author} {\bibfnamefont {T.}~\bibnamefont {Maimbourg}},
  \bibinfo {author} {\bibfnamefont {H.}~\bibnamefont {Punzmann}},\ and\
  \bibinfo {author} {\bibfnamefont {M.}~\bibnamefont {Shats}},\ }\href
  {https://doi.org/10.1103/PhysRevLett.109.114502} {\bibfield  {journal}
  {\bibinfo  {journal} {Physical Review Letters}\ }\textbf {\bibinfo {volume}
  {109}},\ \bibinfo {pages} {114502} (\bibinfo {year} {2012})}\BibitemShut
  {NoStop}%
\bibitem [{\citenamefont {Chen}\ \emph {et~al.}(2018)\citenamefont {Chen},
  \citenamefont {Liu},\ and\ \citenamefont {I}}]{chen_identifying_2018}%
  \BibitemOpen
  \bibfield  {author} {\bibinfo {author} {\bibfnamefont {H.-Y.}\ \bibnamefont
  {Chen}}, \bibinfo {author} {\bibfnamefont {C.-Y.}\ \bibnamefont {Liu}},\ and\
  \bibinfo {author} {\bibfnamefont {L.}~\bibnamefont {I}},\ }\href
  {https://doi.org/10.1103/PhysRevFluids.3.064401} {\bibfield  {journal}
  {\bibinfo  {journal} {Physical Review Fluids}\ }\textbf {\bibinfo {volume}
  {3}},\ \bibinfo {pages} {064401} (\bibinfo {year} {2018})}\BibitemShut
  {NoStop}%
\bibitem [{\citenamefont {Ezersky}\ \emph {et~al.}(1995)\citenamefont
  {Ezersky}, \citenamefont {Ermoshin},\ and\ \citenamefont
  {Kiyashko}}]{ezersky_dynamics_1995}%
  \BibitemOpen
  \bibfield  {author} {\bibinfo {author} {\bibfnamefont {A.~B.}\ \bibnamefont
  {Ezersky}}, \bibinfo {author} {\bibfnamefont {D.~A.}\ \bibnamefont
  {Ermoshin}},\ and\ \bibinfo {author} {\bibfnamefont {S.~V.}\ \bibnamefont
  {Kiyashko}},\ }\href {https://doi.org/10.1103/PhysRevE.51.4411} {\bibfield
  {journal} {\bibinfo  {journal} {Physical Review E}\ }\textbf {\bibinfo
  {volume} {51}},\ \bibinfo {pages} {4411} (\bibinfo {year}
  {1995})}\BibitemShut {NoStop}%
\bibitem [{\citenamefont {Tufillaro}\ \emph {et~al.}(1989)\citenamefont
  {Tufillaro}, \citenamefont {Ramshankar},\ and\ \citenamefont
  {Gollub}}]{tufillaro_order-disorder_1989}%
  \BibitemOpen
  \bibfield  {author} {\bibinfo {author} {\bibfnamefont {N.~B.}\ \bibnamefont
  {Tufillaro}}, \bibinfo {author} {\bibfnamefont {R.}~\bibnamefont
  {Ramshankar}},\ and\ \bibinfo {author} {\bibfnamefont {J.~P.}\ \bibnamefont
  {Gollub}},\ }\href {https://doi.org/10.1103/PhysRevLett.62.422} {\bibfield
  {journal} {\bibinfo  {journal} {Physical Review Letters}\ }\textbf {\bibinfo
  {volume} {62}},\ \bibinfo {pages} {422} (\bibinfo {year} {1989})}\BibitemShut
  {NoStop}%
\bibitem [{\citenamefont {Wright}\ \emph {et~al.}(1996)\citenamefont {Wright},
  \citenamefont {Budakian},\ and\ \citenamefont
  {Putterman}}]{wright_diffusing_1996}%
  \BibitemOpen
  \bibfield  {author} {\bibinfo {author} {\bibfnamefont {W.~B.}\ \bibnamefont
  {Wright}}, \bibinfo {author} {\bibfnamefont {R.}~\bibnamefont {Budakian}},\
  and\ \bibinfo {author} {\bibfnamefont {S.~J.}\ \bibnamefont {Putterman}},\
  }\href {https://doi.org/10.1103/PhysRevLett.76.4528} {\bibfield  {journal}
  {\bibinfo  {journal} {Physical Review Letters}\ }\textbf {\bibinfo {volume}
  {76}},\ \bibinfo {pages} {4528} (\bibinfo {year} {1996})}\BibitemShut
  {NoStop}%
\bibitem [{\citenamefont {Shani}\ \emph {et~al.}(2010)\citenamefont {Shani},
  \citenamefont {Cohen},\ and\ \citenamefont
  {Fineberg}}]{shani_localized_2010}%
  \BibitemOpen
  \bibfield  {author} {\bibinfo {author} {\bibfnamefont {I.}~\bibnamefont
  {Shani}}, \bibinfo {author} {\bibfnamefont {G.}~\bibnamefont {Cohen}},\ and\
  \bibinfo {author} {\bibfnamefont {J.}~\bibnamefont {Fineberg}},\ }\href
  {https://doi.org/10.1103/PhysRevLett.104.184507} {\bibfield  {journal}
  {\bibinfo  {journal} {Physical Review Letters}\ }\textbf {\bibinfo {volume}
  {104}},\ \bibinfo {pages} {184507} (\bibinfo {year} {2010})}\BibitemShut
  {NoStop}%
\bibitem [{\citenamefont {von Kameke}\ \emph {et~al.}(2011)\citenamefont {von
  Kameke}, \citenamefont {Huhn}, \citenamefont {Fernández-García},
  \citenamefont {Muñuzuri},\ and\ \citenamefont
  {Pérez-Muñuzuri}}]{von_kameke_double_2011}%
  \BibitemOpen
  \bibfield  {author} {\bibinfo {author} {\bibfnamefont {A.}~\bibnamefont {von
  Kameke}}, \bibinfo {author} {\bibfnamefont {F.}~\bibnamefont {Huhn}},
  \bibinfo {author} {\bibfnamefont {G.}~\bibnamefont {Fernández-García}},
  \bibinfo {author} {\bibfnamefont {A.~P.}\ \bibnamefont {Muñuzuri}},\ and\
  \bibinfo {author} {\bibfnamefont {V.}~\bibnamefont {Pérez-Muñuzuri}},\
  }\href {https://doi.org/10.1103/PhysRevLett.107.074502} {\bibfield  {journal}
  {\bibinfo  {journal} {Physical Review Letters}\ }\textbf {\bibinfo {volume}
  {107}},\ \bibinfo {pages} {074502} (\bibinfo {year} {2011})}\BibitemShut
  {NoStop}%
\bibitem [{\citenamefont {Francois}\ \emph {et~al.}(2014)\citenamefont
  {Francois}, \citenamefont {Xia}, \citenamefont {Punzmann}, \citenamefont
  {Ramsden},\ and\ \citenamefont {Shats}}]{francois_three-dimensional_2014}%
  \BibitemOpen
  \bibfield  {author} {\bibinfo {author} {\bibfnamefont {N.}~\bibnamefont
  {Francois}}, \bibinfo {author} {\bibfnamefont {H.}~\bibnamefont {Xia}},
  \bibinfo {author} {\bibfnamefont {H.}~\bibnamefont {Punzmann}}, \bibinfo
  {author} {\bibfnamefont {S.}~\bibnamefont {Ramsden}},\ and\ \bibinfo {author}
  {\bibfnamefont {M.}~\bibnamefont {Shats}},\ }\href
  {https://doi.org/10.1103/PhysRevX.4.021021} {\bibfield  {journal} {\bibinfo
  {journal} {Physical Review X}\ }\textbf {\bibinfo {volume} {4}},\ \bibinfo
  {pages} {021021} (\bibinfo {year} {2014})}\BibitemShut {NoStop}%
\bibitem [{\citenamefont {Fauve}\ \emph {et~al.}(1992)\citenamefont {Fauve},
  \citenamefont {Kumar}, \citenamefont {Laroche}, \citenamefont {Beysens},\
  and\ \citenamefont {Garrabos}}]{fauve_parametric_1992}%
  \BibitemOpen
  \bibfield  {author} {\bibinfo {author} {\bibfnamefont {S.}~\bibnamefont
  {Fauve}}, \bibinfo {author} {\bibfnamefont {K.}~\bibnamefont {Kumar}},
  \bibinfo {author} {\bibfnamefont {C.}~\bibnamefont {Laroche}}, \bibinfo
  {author} {\bibfnamefont {D.}~\bibnamefont {Beysens}},\ and\ \bibinfo {author}
  {\bibfnamefont {Y.}~\bibnamefont {Garrabos}},\ }\href
  {https://doi.org/10.1103/PhysRevLett.68.3160} {\bibfield  {journal} {\bibinfo
   {journal} {Phys. Rev. Lett.}\ }\textbf {\bibinfo {volume} {68}},\ \bibinfo
  {pages} {3160} (\bibinfo {year} {1992})},\ \bibinfo {note} {publisher:
  American Physical Society}\BibitemShut {NoStop}%
\bibitem [{\citenamefont {Cross}(1979)}]{cross_relation_1979}%
  \BibitemOpen
  \bibfield  {author} {\bibinfo {author} {\bibfnamefont {M.~M.}\ \bibnamefont
  {Cross}},\ }\href {https://doi.org/10.1007/BF01520357} {\bibfield  {journal}
  {\bibinfo  {journal} {Rheologica Acta}\ }\textbf {\bibinfo {volume} {18}},\
  \bibinfo {pages} {609} (\bibinfo {year} {1979})}\BibitemShut {NoStop}%
\end{thebibliography}%
\bibliographystyle{apsrev4-2}
\end{document}